\documentclass{aa}  
\usepackage{graphicx}

\usepackage{color}

\usepackage{txfonts,ulem}
\usepackage[dvipsnames]{xcolor}

\usepackage{float}
\begin{document}
   \title{A long-term study of the magnetic field and activity in the M giant RZ~Ari\thanks{Based on data obtained using the T\'elescope Bernard Lyot at Observatoire du Pic du Midi, CNRS and Universit\'e de Toulouse, France.}}
  \subtitle{Magnetism and planet engulfment in a fairly evolved star?}

   \author{R. Konstantinova-Antova\inst{1}, S. Georgiev\inst{1,2}, A. L\`ebre\inst{2}, A. Palacios\inst{2}, J. Morin\inst{2}, R. Bogdanovski\inst{1}, C. Abbott\inst{3}, F. Baron\inst{3}, M. Auri\`ere\inst{4}, N. A. Drake\inst{5,6,7}, S. Tsvetkova\inst{1,2}, E. Josselin\inst{2}, C. Paladini\inst{8}, P. Mathias\inst{4}, and R. Zamanov\inst{1}}

   \offprints{R. Konstantinova-Antova}

   \institute{Institute of Astronomy and NAO, Bulgarian Academy of Sciences, 72 Tsarigradsko shose, 1784 Sofia,
             Bulgaria\\
              \email{renada@astro.bas.bg} 
\and
 LUPM, Universit\'e de Montpellier, CNRS,  place Eug\`ene Bataillon, 34095 Montpellier, France             
\and
Department of Physics and Astronomy, Georgia State University, PO Box 5060 Atlanta, Georgia 30302-5060, USA
\and
IRAP, Universit\'e de Toulouse, CNRS, Observatoire Midi Pyr\'en\'es,  57 
Avenue d'Azereix, 65008 Tarbes, France
\and
Laboratory of Observational Astrophysics, Saint Petersburg State University, Universitetski pr. 28, Petrodvoretz 198504, Saint Petersburg, Russia 
\and
Observat\'{o}rio Nacional/MCTI, Rua Gen. Jos\'{e}  Cristino 77, 20921-400, Rio de Janeiro, Brazil
\and
 Laborat\'{o}rio Nacional de Astrof\'{\i}sica/MCTI, Rua dos Estados Unidos 154, 377504-364, Itajub\'{a}, Brazil 
\and
ESO — European Southern Observatory Alonso de Córdova 3107 Santiago, Chile 
}

   \date{Received ; accepted }

\abstract 
{}
{We present a detailed long-term study of the single M6 III giant RZ~Ari to obtain direct and simultaneous 
measurements of the magnetic field, activity indicators, and radial velocity in 
order to infer the origin of its activity. We study its magnetic activity in the context of stellar evolution, and for this purpose, we also refined its evolutionary status and Li abundance. In general, for the M giants, little is known about the properties of the magnetic activity and its causes. RZ~Ari posses the strongest surface magnetic field of the known Zeeman-detected M giants and is bright enough to allow a deep study of its surface magnetic structure. The results are expected to shed light on the activity mechanism in these stars.}
{We used the spectropolarimeter Narval  at the \textit{T{\'e}lescope Bernard Lyot}
(\textit{Observatoire du Pic du Midi}, France) to obtain a series of Stokes~$I$ and 
Stokes~$V$ profiles for RZ~Ari. Using the least-squares deconvolution technique, we were able to 
detect the Zeeman signature of the magnetic field. We measured its 
longitudinal component by means of the averaged Stokes~$V$ and Stokes~$I$ 
profiles. In addition, we also applied Zeeman-Doppler imaging (ZDI) to search for the rotation period of the star, and we constructed a tentative magnetic map. It is the first magnetic map for a star that evolved at the tip of red giant branch (RGB) or even on the  asymptotic giant branch (AGB). The spectra also allowed us to monitor chromospheric emission lines, which are well-known indicators of stellar magnetic activity.
From the observations obtained between September 2010 and August 2019,
we studied the variability of the magnetic field of RZ~Ari. We also redetermined the initial 
mass and evolutionary status of this star based on current stellar evolutionary 
tracks and on the angular diameter measured from CHARA interferometry.} 
{Our results point to an initial mass of 1.5~M$_{\odot}$, so that this giant
is more likely an early-AGB star, but a lotaction at the tip of the RGB is not completely excluded. With a 
$v\sin i$ of
6.0 
$\pm$ 0.5~km\,s$^{-1}$, the upper limit for the rotation period is found to be 909 days. On the basis of our dataset and AAVSO photometric data, we determined periods longer than 1100 days for the magnetic field and photometric variability, and 704 days for the spectral line activity indicators. The rotation period determined on the basis of the Stokes~$V$ profiles variability is 530 days. A similar period of 544 days is also found for the photometric data. When we take this rotation period and the convective turnover time into account, an effective action of an $\alpha$-$\omega$-type dynamo seems to be unlikely, but other types of dynamo could be operating there. The star appears to lie outside the two magnetic strips on the giant branches, where the $\alpha$-$\omega$-type dynamo is expected to operate effectively, 
and it also has a much higher lithium content than the evolutionary model predicts. These facts suggest that a planet engulfment could speed up its rotation and trigger dynamo-driven magnetic activity. On the other hand, the period of more than 1100 days  cannot be explained by rotational modulation and could be explained by the lifetime of large convective structures. The absence of linear polarization at the time the magnetic field was detected, however, suggests that a local dynamo probably does not contribute significantly to the magnetic field, at least for that time interval.
}
{}{}

   \authorrunning {Konstantinova-Antova et al.}
   \titlerunning {A long-term study of the magnetic field and activity in the M giant RZ~Ari}
   \maketitle 


\section{Introduction}
Recently, magnetic fields were detected in many single G, K, and M giants \citep{KA2010,KA2013,KA2014,Auriere2015}. While the main origins of the magnetic fields in G and K giants are the $\alpha$-$\omega$ dynamo and a remnant magnetic field in the Ap star descendants, in the case of M giants, the origins of the magnetic field are not completely clear. \cite{Charbonnel2017} considered that the $\alpha$-$\omega$ dynamo might operate even in early asymptotic giant branch (AGB) stars due to the properties of their convective envelopes. In addition, some of these stars possess fast rotation that cannot yet be explained by the theory of stellar evolution (Zamanov et al. 2008; Konstantinova-Antova et al. 2010, 2013). On the other hand, most of the M giants are pulsating, and it is still unclear how their pulsations might relate to the magnetic field generation. For example, the magnetic M giants EK~Boo and $\beta$ Peg we studied \citep{Georgiev2020betpeg} are known as semiregular variable stars (Samus et al. 2017; Tabur et al. 2009).

\noindent It was also recently found that the magnetic M giants occupy a certain area in the Hertzsprung-Russel diagram (HRD), the so-called second magnetic strip \citep{KA2014}. This strip coincides with the tip of the red giant branch (RGB) and early AGB phase, in agreement with the theoretical models of \cite{Charbonnel2017}. 

\noindent RZ~Arietis = HD~18191 is a 6 V magitude single star of spectral class M6~III,  according to Simbad at the CDS. It has the strongest longitudinal magnetic field ($|B_l|_{\rm max}=14$~G) of all the M giants we studied  \citep{KA2013,KA2014}, and it has a relatively large $v\sin i$ 6.0 km\,s$^{-1}$ \citep{Georgiev2020}. Moreover, the star is bright enough to allow a deep study of its surface
magnetic structure. According to \cite{KA2010}, its effective temperature $T_{\rm eff}$ is 3450~K and $\log(L/L_{\odot}$) is 2.75. In these previous studies, we found that the star has 1.0~M$_{\odot}$ and is located either near the RGB tip or at the beginning of the AGB. A detailed study of the properties of the magnetism in this star is expected to provide knowledge about the mechanism that generates the magnetic field at these evolutionary stages when the stars develop a complex structure through dynamos of different types (interface and/or a local dynamo). 

\noindent In addition, RZ~Ari is also known as a semiregular variable star with a pulsation period of about 50 days and a long secondary period (LSP) of 480 days \citep{Percy2008,Percy2016,Tabur2009}. It also has a long-term photometric light curve according to AAVSO \footnote{The American Association of Variable Star Observers:\newline https://www.aavso.org/} observations. All these data allow us to test whether the eventual shock waves during pulsations play a role in its magnetic field generation, as was found for some Mira-type and RV Tau stars (Lebre et al. 2014; Georgiev et al. 2023).

\noindent We present the results of about ten years of spectropolarimetric observations and magnetic activity study for this fairly evolved giant with relatively fast rotation, and we try to explain the origin of its magnetic field. A deep study like this has never been conducted before for RZ Ari or another magnetic M giant at the tip of the RGB and AGB evolutionary stage. The observations and data measurements are presented in Section 2. Section 3 contains the determination of the fundamental parameters and evolutionary status of the star. The results for its magnetic field and activity are described in Section 4. Section 5 describes our observations of the linear polarization of the star. Section 6 contains a discussion of the origins for the magnetic field in this M giant, and Section 7 gives our conclusions. The journal of the observations, together with the measurements of the activity indicators and longitudinal magnetic field, is presented in Appendix A. The first tentative magnetic map for a star at this stage of evolution is presented in Appendix B.


\begin{figure}[t]
    \centering
    \includegraphics[width=\columnwidth]{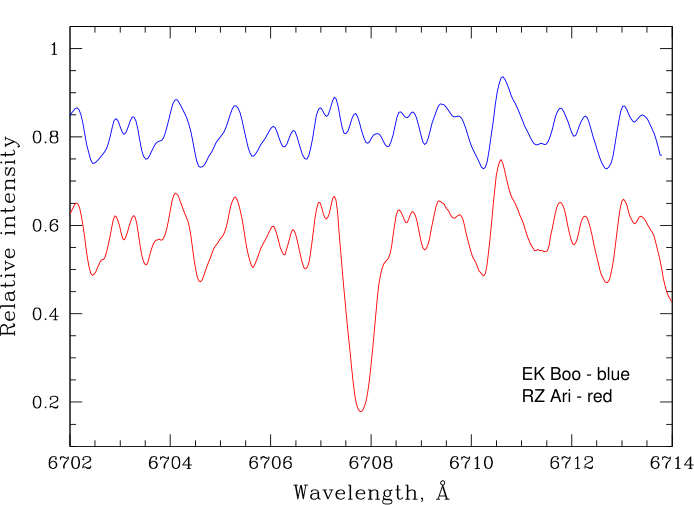}
    \caption{Observed spectrum of  RZ~Ari around the \ion{Li}{i} 6707.8~\AA\ resonance line. The spectrum of EK~Boo in the same region is also plotted for comparison (blue line) and is shifted in relative intensity by +0.2 for clarity.} 
    \label{fig:rzari_ekboo}
\end{figure}

\section{Observations and data processing}
\label{sec:observations}

The observations of RZ~Ari were carried out at the 2 m \textit{T{\'e}lescope Bernard Lyot} (TBL) of the \textit{Pic du Midi} observatory, with the spectropolarimeter Narval \citep{Auriere2003}. Narval is 
a copy of the instrument ESPaDOnS, in operation at the 3.60m CFHT \citep{Donati2006}. It is a fiber-fed 
\textit{{\'e}chelle} spectrometer, able to cover the whole spectrum from 370 nm to 1000 nm 
in a single exposure. Forty orders are aligned on the CCD frame, 
separated by two cross-disperser prisms. We used Narval in polarimetric 
mode, with a spectral resolution of 65\,000. Stokes $I$ (unpolarized) and 
Stokes $V$ (circular polarization) parameters were measured. Each Narval exposure consists of
four subexposures, between which the retarders (Fresnel rhombs) are rotated 
in order to exchange the beams in the instrument and to reduce spurious 
polarization signatures. 

\noindent We observed RZ~Ari in the period September 2010 -- August 2019. We performed series of several exposures per night (see Appendix A). They were averaged after the processing.
To extract the spectra, we used \textit{Libre-ESpRIT} \citep{Donati1997}, a fully automatic reduction package installed at 
TBL. For the Zeeman analysis, a least-squares deconvolution (LSD) 
\citep[LSD,][]{Donati1997} was applied to all spectra. 
In the LSD process, we used a mask calculated for solar chemical abundances, an effective 
temperature of 3400~K, $\log g =0.5$, and a microturbulence velocity of 2.0 km\,s$^{-1}$. 
These parameters are consistent with the spectral class and luminosity of 
RZ~Ari \citep{KA2010}. This method enabled us to average a total of about 
12 000 atomic spectral lines and obtain their mean Stokes $I$ and Stokes $V$ profiles.  
The null spectrum ($N$ profile) given by the standard procedure \citep{Donati1997} was also examined, but never showed any 
signal. This confirms that the detected signatures in the V profiles are not spurious. From the obtained mean Stokes~$V$ for each night, we computed the 
surface-averaged longitudinal magnetic field $B_{l}$ in G, using the 
first-order moment method \citep{Donati1997,RS1979}.\\

\noindent The LSD profiles obtained with this classical line mask were calculated by taking approximately 12~000 lines into account that are formed at different altitudes in the extended atmosphere of the star. In order to focus on the lines that represent the physical conditions at the bottom of the atmosphere, we created another LSD line mask by taking the classical mask and restricting the lines used on the basis of their excitation potential, as was done for the post-AGB star R~Sct by \cite{Georgiev2023}. In general, the higher the excitation potential of an atomic line, the narrower its layer of formation in the stellar atmosphere. Lines with high excitation potential are mostly formed in the lower parts of the atmosphere, that is, close to the photosphere. On the other hand, lines with low excitation potential may be formed at more extended layers in the atmosphere, and thus their profiles give a mixed representation of the conditions at both lower (near-photospheric) and higher atmospheric layers. This means that if we desire LSD profiles that mostly represent the conditions at the bottom of the atmosphere, we must use a mask that consists of lines with a sufficiently high excitation potential, that is, higher than some threshold value $\chi_{\rm thr}$. However, we must also keep $\chi_{\rm thr}$ low enough so that the number of lines that remain in the mask is sufficient to allow the calculation of an LSD profile with a significant signal-to-noise ratio (S/N). After experimenting with different values, we found that $\chi_{\rm thr} = 2.5$~eV is both high enough to filter out the contribution of the higher atmosphere and low enough to allow a sufficient number of lines (4851) into the line mask. The LSD Stokes~$I$ profiles of RZ~Ari obtained with masks of $\chi_{\rm thr} = 2.5$~eV are almost the same as those obtained with higher $\chi_{\rm thr}$, which shows that at 2.5~eV, the bottom layers of the atmosphere are essentially reached.\\\\
Because the LSD profiles computed from lines with high excitation potential represent the conditions at the bottom of the atmosphere, they should in principle be ideal for studying the surface magnetic field through the LSD Stokes~$V$ profile and the longitudinal magnetic field, $B_l$, which is calculated from that same profile. However, unfortunately, the decreased number of lines in the mask drastically deteriorates the S/N of the Stokes~$V$ profiles, leading to a loss of precision and a very large error bar in the $B_l$ computations. We therefore used the restricted LSD line mask for the analysis of the stellar radial velocity at the level of the photosphere (estimated by fitting the LSD Stokes~$I$ profiles with a Gaussian) in Section~\ref{sec:bl_ind_vrad}, and the full LSD line mask for the calculation of the $B_l$ values in Section~\ref{sec:bl_ind_vrad} and the analysis of the Stokes~$V$ profiles via the Zeeman-Doppler imaging (ZDI) method in Section~\ref{sec:period-zdi}.

\noindent The magnetic activity of RZ~Ari was monitored by means 
of measurements of the relevant indexes of the chromospheric activity indicators Ca\,{\sc ii}~K, and H, H$\alpha$, and the calcium infrared triplet (Ca IRT). Appendix A lists the journal of our observations of RZ~Ari, including the dates, the heliocentric Julian day (HJD), the total exposure time, the detection level from the LSD statistics \citep{Donati1997}, $B_{l}$ and its error (in G), the activity indicator indexes, and the radial velocity.


\section{Fundamental parameters and evolutionary status of RZ~Ari}

\subsection{Radius, projected rotational velocity, and rotation period}

Taking into account
the angular diameter of RZ~Ari of 10.22 mas measured during lunar occultation \citep{Richichi2006} and the distance
to the star 97.7794 pc as provided by the Gaia DR2 catalog \citep{GaiaDR2}, we obtain a radius of 107.43 R$_{\odot}$ that is consistent with both the RGB tip and the early-AGB phase. 
In 2020 we obtained new angular diameter measurements with the CHARA interferometer for RZ~Ari. The data were gathered with the MIRC-X instrument \citep{Kraus2018,Anugu2018} using all six telescopes of the array. The reduction and calibration were performed using the MIRC-X pipeline \citep{Lebouquin2020}.  The angular size, limb-darkening coefficients, and corresponding errors were determined from these data with model-fitting and bootstrap methods using the OITOOLS library of interferometric analysis and imaging techniques \citep{Baron2019}. The determined angular diameter is 10.27 $\pm$ 0.007 mas, which agrees well with the previous measurements, and results in a radius of about 107.9$\pm$ 6.2 R$_{\odot}$. Using the temperature of $3400 \pm 100$~K \citep{Georgiev2020}, we find a luminosity $\log(L/L_\odot) = 3.144$ \footnote{We have used here L$_\odot = 3,846 \times 10^{33}$ in cgs units, for consistency reasons with stellar evolution models.}, which is fully consistent with the value of 3.15 found by \cite{Villaume2017} from IR measurements. 

\noindent A projected rotational velocity $v\sin i$ of 9 km\,s$^{-1}$ was determined for the first time by \cite{Zamanov2008}. This was later revised to 6.0 $\pm$ 0.5km\,s$^{-1}$ by \cite{Georgiev2020}, taking the macroturbulence of the star into account. With this value and the radius of about 108 R$_\odot$, the upper limit for the period P is about 909 days.

\begin{figure}[t] 
    \centering
    \includegraphics[width=\columnwidth]{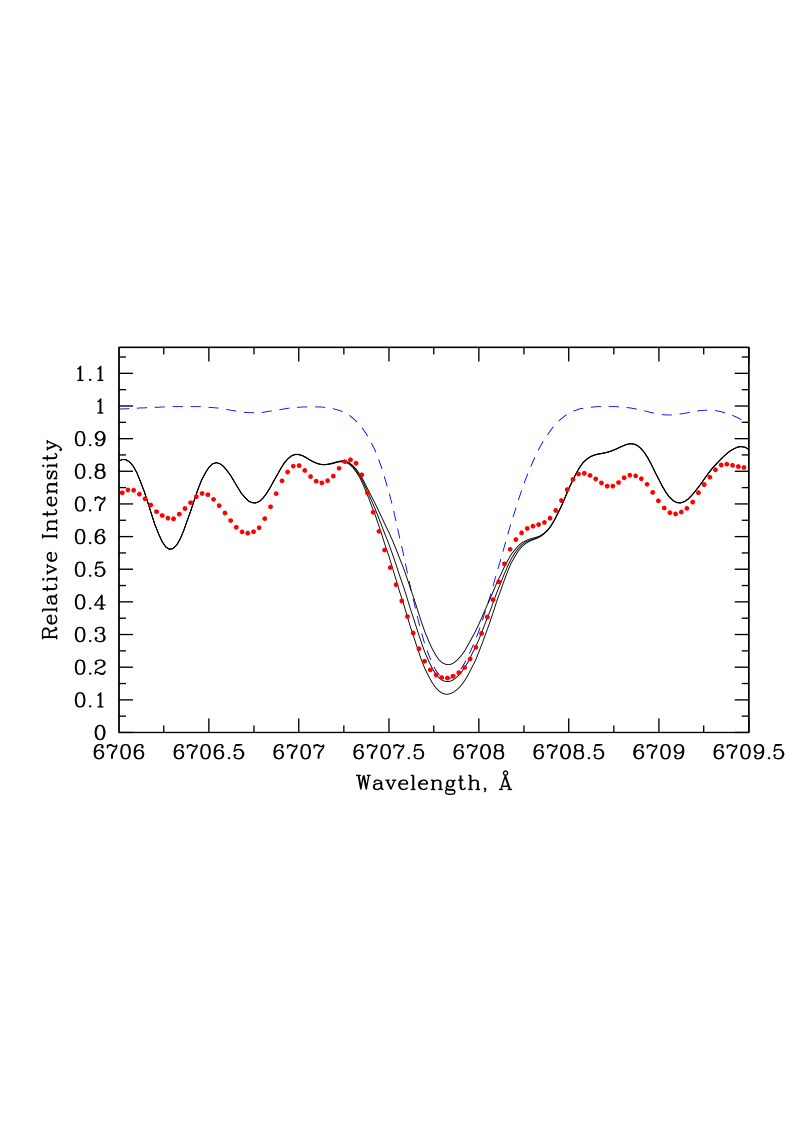}
    \caption{Observed (red dots) and synthetic (solid black lines) spectra of RZ~Ari in the region around the \ion{Li}{i} 6707.8~\AA  resonance line. The synthetic spectra were calculated with lithium abundances of $\log\varepsilon({\rm Li})$ = 0.8, 1.2, and 1.6. The dotted blue line is the synthetic spectrum calculated without the contribution of the TiO molecular lines. The best fit is achieved for the  $\log\varepsilon({\rm Li})$= 1.2 $\pm 0.2$ dex.}
    \label{fig:rzari-Lithium}
\end{figure}

\subsection{Lithium abundance}  
\label{sec:spectrum}

A very interesting characteristic of the RZ~Ari spectrum is the strong resonance \ion{Li}{i} line at $\lambda$6707.8~\AA. In Figure~\ref{fig:rzari_ekboo} we compare the spectrum of RZ~Ari with the spectrum of another late-M giant, EK~Boo (HD~130144), that was studied in \cite{KA2010}. Both stars have very similar atmospheric parameters, such as $T_{\rm eff}$ and $\log  g$. Moreover, magnetic fields have been detected in both stars. As shown in \cite{KA2010} for EK~Boo, there is no noticeable presence of the line \ion{Li}{i} 6707.8~\AA. By means of synthetic spectra calculation, the upper limit to the lithium abundance was determined as $\log\varepsilon({\rm Li}) \le -0.8$. 

\noindent In order to determine the Li abundance in RZ~Ari, we used the wavelengths and oscillator strength for the individual hyperfine and isotopic components of the lithium resonance line from \cite{hobbs1999}. The list for the $^{48}$TiO lines, which are very strong in this spectral region, was taken from \cite{plez1998}. The details of the atomic and molecular lines parameters are given in \cite{KA2010}. The atmospheric parameters of RZ~Ari were estimated from the Gaia DR2 distance, angular diameter, and $T_{\rm eff}$ determination (see section~3). In the synthetic spectra calculations, we used the grid of MARCS models, computed with carbon and nitrogen abundances moderately changed by the first dredge-up, as expected for the atmospheres of M-type giants. The best fit, shown in Fig.~\ref{fig:rzari-Lithium}, is achieved for the $\log\varepsilon({\rm Li})=1.2 \pm 0.2$. This value is compatible with a post-first-dredge-up abundance as predicted from standard stellar
evolution models. However, the seminal work of \cite{Brown1989} showed that only about 1 \% of the observed K giant stars that are ascending the RGB are Li-rich, and with the exception of these stars, the average abundance of Li in the their atmospheres is about $\log\varepsilon({\rm Li})~\approx$~+0.1, that is to say, much lower than the abundance we derive for RZ-Ari. More recently, \cite{Charbonnel2020} reanalyzed a large sample of red giant stars with Gaia DR2 parallaxes and determined their lithium abundances. This sample included a few dozen M giants, for which the lithium abundance is indeed always lower than +0.5. Within this framework, the lithium abundance we determined in RZ~Ari can be considered as higher than that of most M giants with a known lithium abundance.\\

\begin{table}[t]
    \caption{Chemical composition  and stellar parameters of RZ~Ari as derived in this work and found in previous spectroscopic studies. The adopted fundamental parameters are marked in boldface.}
    \centering
    \begin{tabular}{cc c}
    \hline \hline 
      Parameter & Abundance   & Source \\
      \hline 
        A($^7$Li) & 1.2 $\pm 0.2$ & This work\\
        & & \\
        $^{12}$C/$^{13}$C &  6 $\pm 1$ & \cite{Lebzelter2019}\\
        & 7.9 $\pm$ 0.8 & \cite{Tsuji2008} \\
        & & \\
        $^{16}$O/$^{17}$O & 958$^{+912}_{-467}$ &  \cite{Lebzelter2019}\\
        & 607$\pm 48$ & \cite{Tsuji2008}\\
        & & \\
        $^{16}$O/$^{18}$O  & 2422$^{+1185}_{-796}$& \cite{Lebzelter2019} \\
        & & \\
        $\rm{[Fe/H]}$ & \bf{-0.25} $\pm$ 0.12 & \cite{Prugniel2011}\\
        \hline
        & & \\
       Mass (M$_\odot$) & \bf{1.5 $\pm 0.4$} & \cite{Tsuji2008}\\
       & & \\
       log(L /L$_\odot$) & 3.178 & \cite{Lebzelter2019}\\
       & 3.15 & \cite{Villaume2017} \\
       & \bf{3.144} & This work\\
       & & \\
       log g & 0.3 $\pm$ 0.26 & \cite{Prugniel2011} \\
        & 0.36 & \cite{Lebzelter2019} \\
       & & \\
       R (R$_\odot$) & 123.728 & \cite{Lebzelter2019}\\
       & \bf{107.91 $\pm$ 6.2 } & This work \\
       & & \\
       T$_{\rm eff}$ (K) & 3236 &  \cite{Lebzelter2019} \\
       & 3250 $\pm 38$ & \cite{Prugniel2011}\\
       & 3341 & \cite{Tsuji2008} \\
       & 3442 $\pm$ 148 & \cite{Dyck1998} \\
       & {\bf 3400 $\pm$ 100} & \cite{Georgiev2020} \\
      & & \\
       $\upsilon \sin i$ (km\,s$^{-1}$) & \bf{6.0 $\pm$ 0.5} & \cite{Georgiev2020}\\
       \hline 
    \end{tabular}
    \label{tab:rzari_params}
\end{table}

\subsection{Evolutionary status} \label{sec:evol}

\subsubsection{Chemical composition as evolutionary evidence}
Several abundance determinations exist for RZ Ari, in particular, of the isotopic ratios, which may be very informative about the initial mass and evolutionary status of the star. We retained the Li abundance and the $^{12}{\rm C}/^{13}{\rm C}$, $^{16}{\rm O}/^{17}{\rm O}$ and $^{16}{\rm O}/^{18}{\rm O}$ isotopic ratios as guides. A summary of the values found in the literature for these abundances is given in Table~\ref{tab:rzari_params}.
As observed in most of the evolved low-mass giants \citep{CC1998}, the carbon isotopic ratio is close to nuclear equilibrium value in this star. This indicates that it has undergone additional mixing during the RGB ascent that connected the convective envelope to the hydrogen-burning shell, where hydrogen fusion occurs through the full CNO cycle. These transport processes dig deep into nuclear burning regions and are expected to efficiently destroy lithium \citep{CZ2007,Lagarde2012}. This may represent a challenge given that RZ Ari exhibits a strong line of Li \citep{Merchant1967}, with a lithium abundance of 1.2 $\pm 0.2$ dex (this work; see Section~\ref{sec:spectrum}) typical of the classical (without additional mixing) post-first-dregde-up (hereafter FDU) evolution of low-mass stars.
\begin{figure*}
    \centering
    \includegraphics[width=0.5\textwidth]{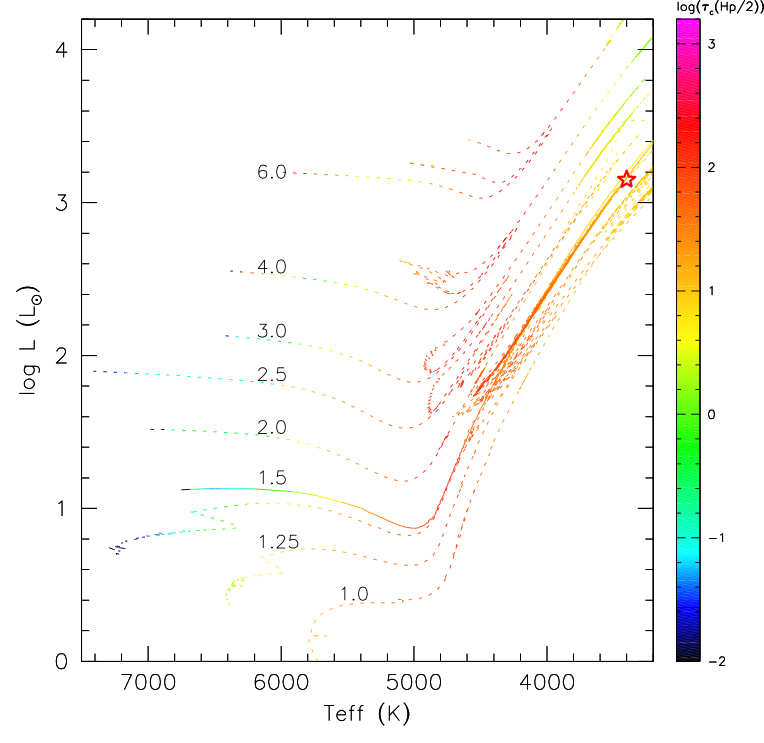}%
    \includegraphics[width=0.47\textwidth]{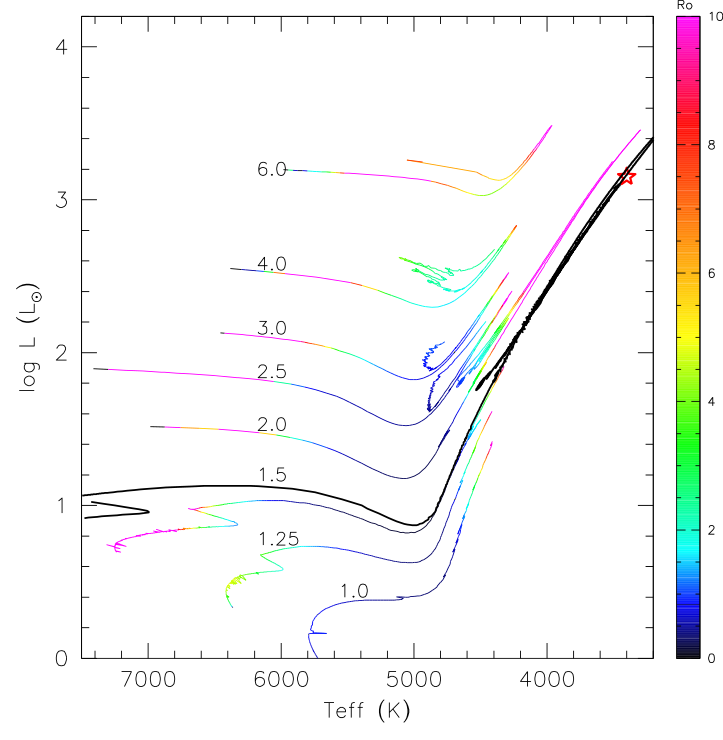}
    \caption{HR diagrams with computed stellar evolutionary tracks of stars of different initial masses. {\em Left panel:} Color-coded convective turnover time $\tau(H_p/2)$ (in logarithmic scale) in the stellar convective envelope along the evolution in the Hertzsprung-Russell diagram for solar metallicity with rotation and thermohaline mixing from \cite{Charbonnel2017} as dashed lines (initial masses of 1, 1.25, 1.5, 2, 3, 4, and 6 M$_\odot$) and our 1.5 M$_\odot$ model at [Fe/H] = -0.25 dex with thermohaline mixing and without rotation (solid line). The red star marks the location of RZ Ari in the HR diagram (see Tab.~\ref{tab:rzari_models}). {\em Right panel:} Same, with the color-coded Rossby number (in linear scale) at half a pressure scale height above the base of the convective envelope for the models from \cite{Charbonnel2017}. The track of the dedicated 1.5 M$_\odot$ nonrotating model at [Fe/H] = -0.25 dex is also shown as the solid black line. A star like this should have the same Rossby number evolution as the colour-coded 1.5 M$_\odot$ track from \cite{Charbonnel2017} when it rotates, e.g., Ro $\approx 10$ at the location of RZ Ari.}
    \label{fig:HRD_RZAri}
\end{figure*}
The oxygen isotopic ratio $^{16}{\rm O}/^{17}{\rm O}$ estimated from molecular bands in RZ Ari is lower than 1000. According to predictions of classical evolution for FDU abundances of cool giants, these values are reached by models with initial masses between $1.3 M_\odot \leq M \leq 2 M_\odot$ \citep{Dearborn1992,KarakasLattanzio2014}. The $^{16}{\rm O}/^{17}{\rm O}$ isotopic ratio is mainly modified during the FDU episode and is very sensitive to the initial mass as well as to the adopted nuclear reactions rate for the CNO cycle \citep{Halabi2016}, but it is not significantly affected by the additional mixing process that cause the decrease in the carbon isotopic ratio on the RGB \citep{CL2010}. The $^{16}{\rm O}/^{18}{\rm O}$ isotopic ratio as measured by \cite{Lebzelter2019} exceeds 2000, which is expected for massive AGB stars ($\geq 6 {\rm M}_\odot$) experiencing hot bottom burning (hereafter HBB). However, the $^{16}{\rm O}/^{17}{\rm O}$ isotopic ratio and the luminosity of RZ Ari are incompatible with this evolutionary status, and \cite{Lebzelter2019} argued that it cannot be more massive than 3 M$_\odot$.\\
To summarize, based on its surface chemical composition, RZ~Ari should be the descendant of a star with 1.3 to 2 M$_\odot$. This is compatible with previous mass estimates for this star by means of stellar evolution models \citep[see][]{Tsuji2008,Halabi2016,Lebzelter2019}.
\begin{table}[t]
    \caption{Parameters and chemical composition of RZ~Ari from our model predictions at two evolutionary stages (RGB tip and early AGB) compatible with the radius as derived in this work by interferometry and luminosity as derived from IR measurements by \cite{Villaume2017} (both values highlighted in boldface in Table~\ref{tab:rzari_params}).  The mass is given here at the evolutionary point.}
    \centering
    \begin{tabular}{c c c }
    \hline \hline
        Parameter & RGB & eAGB \\
        \hline
         Mass (M$_\odot$) & 1.37  &  1.21   \\
       & & \\
        log(L /L$_\odot$) & 3.148 &  3.139 \\
       & & \\
         R (R$_\odot$) & 107.92 & 107.92\\
       & & \\
         T$_{\rm eff}$ (K)& 3434 & 3436\\ 
       & & \\
         A($^7$Li) & 0.637 & 0.134\\
       & & \\
         $^{12}$C/$^{13}$C & 17 & 14 \\
       & & \\
         $^{16}$O/$^{17}$O & 688& 655\\
       & & \\
         $^{16}$O/$^{18}$O & 675& 700\\
       & & \\
         $\tau_c$ (days)& 5.48& 11.6\\
         \hline
    \end{tabular}
    \label{tab:rzari_models}
\end{table}

\subsubsection{Stellar evolution models}

 Considering the clues at hand ($T_{\rm {eff}}$, $\log g$, $R_\star$, $L_\star$, and surface chemical abundances), we 
used an updated version of the STAREVOL code \citep[V3.40, see][]{Dumont2021b,Dumont2021a} to compute a dedicated 1.5 M$_\odot$ stellar evolution model and also used the models from \cite{Charbonnel2017}. In contrast to \cite{Lagarde2012}, we adopted the analytical formula of \cite{KS66} for the atmosphere, with a photosphere defined as the layer for which the optical depth $\tau$ is between 0.005 and 10, and we classically defined the effective temperature and radius at $\tau = 2/3$. This choice leads to more realistic tracks on the RGB phase compared to what can be obtained when using a gray atmosphere \citep{Salaris2002}. Convective mixing was modeled according to the mixing length theory with  a parameter $\alpha_{MLT} = 1.6$ as in \cite{Lagarde2012,Charbonnel2017}. This value is lower than that deduced from solar calibration for the setup of physics\footnote{In particular, the treatment of the atmosphere} considered \citep{Dumont2021a}. We modeled mass loss from the ZAMS to the central helium exhaustion according to \cite{Reimers1975}, 
\[\ \dot{M} = -3.98~10^{-13} \eta_{R} \frac{LR}{M} ~M_{\odot}\mathrm{yr^{-1}},\] with a parameter $\eta_R = 0.5$ as in \cite{Lagarde2012}.  Following the results of \cite{Lagarde2012}, we neglected rotation, but took thermohaline mixing into account as prescribed by \cite{CZ2007}. This specific form of double-diffusive convection instability becomes efficient beyond the RGB luminosity bump and may lead, as shown by \cite{Lagarde2012}, to an additional depletion of lithium and a decrease in the carbon isotopic ratio at the surface of evolved RGB stars, in good agreement with observations \citep{Lagarde2019}. We adopted a subsolar ([Fe/H] = -0.25 dex\footnote{According to our spectral analysis and to the value derived by \cite{Prugniel2011}.}) metallicity with a solar scaled mixture of elements based on \cite{AGSS09}. We also adopted the solar isotopic ratios recommended in that work. 
Finally, we adopted the nominal nuclear reaction rates from the Nacre II compilation \citep{NacreII}.

   \begin{figure*}[t]
   \centering
   \includegraphics[width=\textwidth, angle=0]{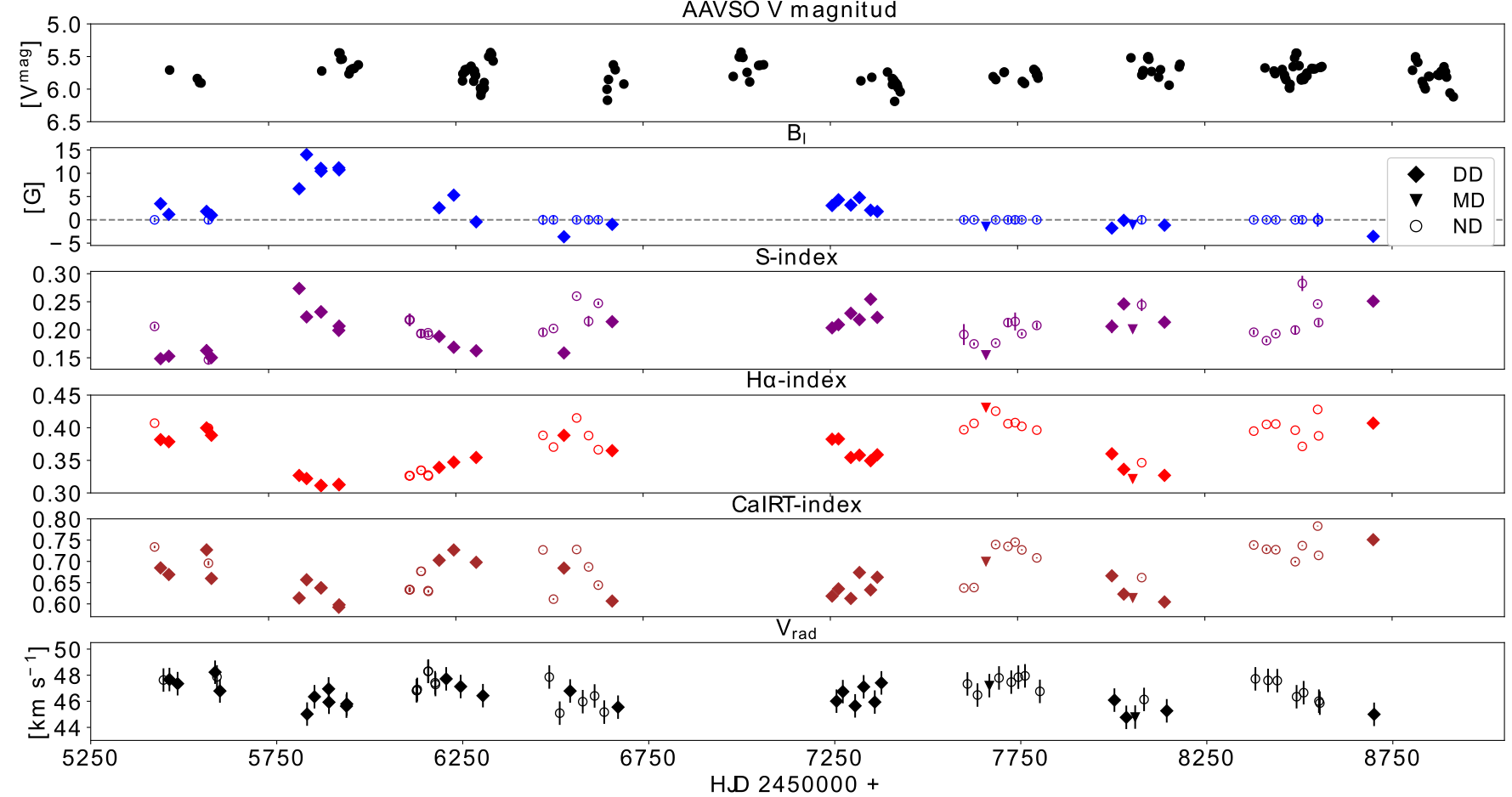}
     \caption{From top to bottom: AAVSO light curve in the V band, longitudinal magnetic field (calculated with the full line mask), activity indicators, and $v_{\rm rad}$ (calculated with the restricted line mask) variability in RZ~Ari for the dates listed
in Table~1. The magnetic field nondetections are designated with open symbols.}
         \label{fig:bl_ind_vrad}
   \end{figure*}

\noindent The predictions of the model are presented in Tab.~\ref{tab:rzari_models} and Fig.~\ref{fig:HRD_RZAri}.\\
Based on our interferometric radius, we are able to find stellar evolution models that comply with the fundamental parameters (luminosity, effective temperature, and metallicity) of RZ~Ari. RZ~Ari is well fit by our 1.5 M$_\odot$ , [Fe/H] = -0.25 dex stellar evolution model. Its location in the HRD does not allow us to determine its evolutionary status as it is both compatible with the RGB tip and the early AGB, and the surface chemical composition is too uncertain (in particular, the oxygen isotopic ratios) for us to be able to clearly distinguish these evolutionary points. The models we used predict very low Li abundances at the RGB tip and the early AGB as a consequence of thermohaline mixing along the RGB ascent (needed to approach the observed carbon isotopic ratio), as Table shows.~\ref{tab:rzari_models}. These values agree with observations for $\approx$ 98\% of the field red giants \citep{Lagarde2012,Charbonnel2020}.
The surface lithium abundance of $\log\varepsilon({\rm Li})$ = 1.2 $\pm$ 0.2 indicates that RZ~Ari  is an actual Li-rich star, as we discussed in \S~\ref{sec:spectrum}. This can be of some importance when we try to understand our observational set (see \S~\ref{sec:discussion}). The location of the giant branches for our model is the same as in the model of 1.5 M$_\odot$ and for the solar metallicity in \cite{Charbonnel2017}, with similar convective turnover timescales at the evolutionary points recorded in Tab.~\ref{tab:rzari_models} as well. For this reason, and even though rotation and rotational mixing are not included in our model, we can refer to that work to estimate a theoretical Rossby number, which should be greater than 10 (and about 100; see Fig.~\ref{fig:HRD_RZAri}). It lies beyond and well above the magnetic strips defined in \cite{Auriere2015} and \cite{KA2014} that were interpreted in \cite{Charbonnel2017} as regions in which the Rossby number drops to or below unity on the RGB, during core He burning and in the earlier AGB phase (dark blue regions in the right panel of Fig.~\ref{fig:HRD_RZAri}), permitting the development of an $\alpha - \omega$ dynamo. 

\section{Magnetic field and activity in RZ~Ari}

The magnetic field and activity in RZ~Ari were monitored over the period 2010-2019 with the spectropolarimeter Narval. The log of observations and the measured longitudinal magnetic field $B_l$, the radial velocity, and the activity indicators (S-index, $H_\alpha$ index, and Ca IRT index) are presented in Table~\ref{tab:rzari_v}. Their variability together with the photometric variability by AAVSO data is shown in Fig.~\ref{fig:bl_ind_vrad}. Table~\ref{tab:rzari_v} and Fig.~\ref{fig:bl_ind_vrad} show that only 27 of the 56 observations show a magnetic field detection. There is a tendency for the magnetic activity to decline after 2015. This might be a result of a long-term magnetic cycle. The measured longitudinal magnetic field between September and November 2011 represents the largest field ever measured in an M-giant star \citep{KA2013,KA2014}. This no longer appears during the monitored period, and the detected longitudinal field has a magnitude 5 G and lower for 20 out of the 27 dates with definite or marginal detections. Periods of detection and nondetection are found also for other long-term studied M giants such as EK~Boo and $\beta$~Peg \citep{Georgiev2020betpeg}, in contrast to the G and K giants near the base of the RGB.

\subsection{Magnetic field, activity indicators, and radial velocity variability of RZ~Ari}
\label{sec:bl_ind_vrad}
The magnetic field and activity indicators show both short and long-term variability.
The comparison of $B_l$ and the activity indicator behavior reveals a lack of strict similarity between the two. In particular, the activity indicators vary even during periods of no variability and no detection of the magnetic field. In these periods, however, the activity indicators show a lower value than in periods of magnetic field detection. Hence, the magnetic field nondetection cannot be explained by a cancellation of fields with opposite polarities.

\noindent In addition, we therefore deduce that the chromospheric heating is not fully magnetic, and that other processes also play a role in the variability of the different indexes presented here.\\

\begin{figure}[H]
    \centering
    \includegraphics[width=0.48\textwidth]{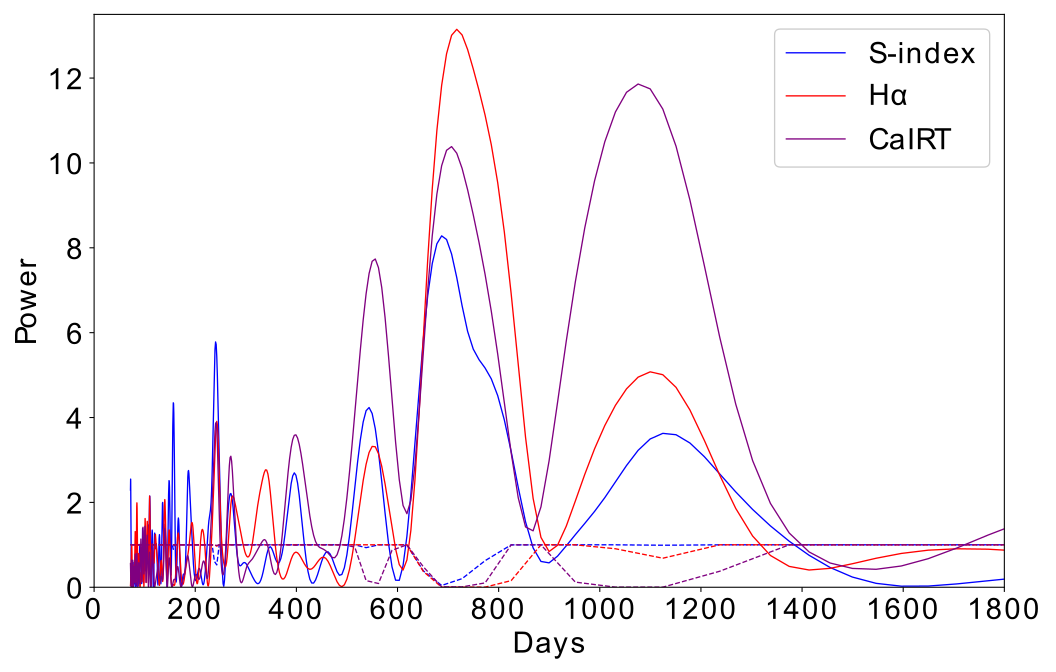}
    \includegraphics[width=0.48\textwidth]{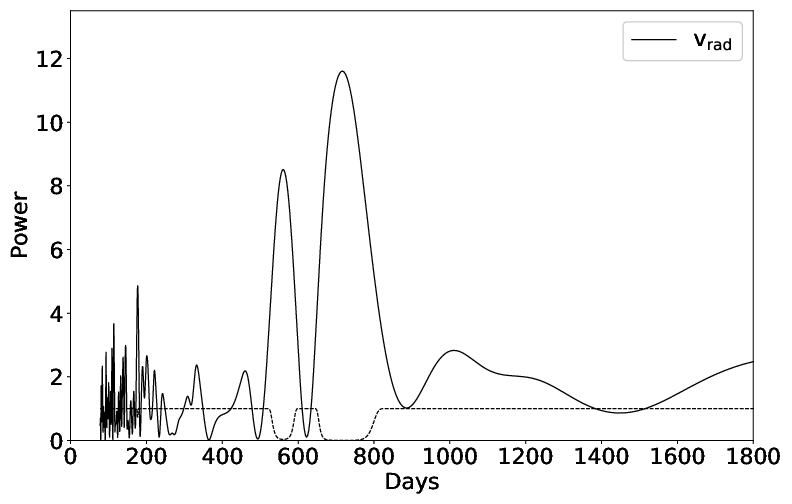}
    \includegraphics[width=0.48\textwidth]{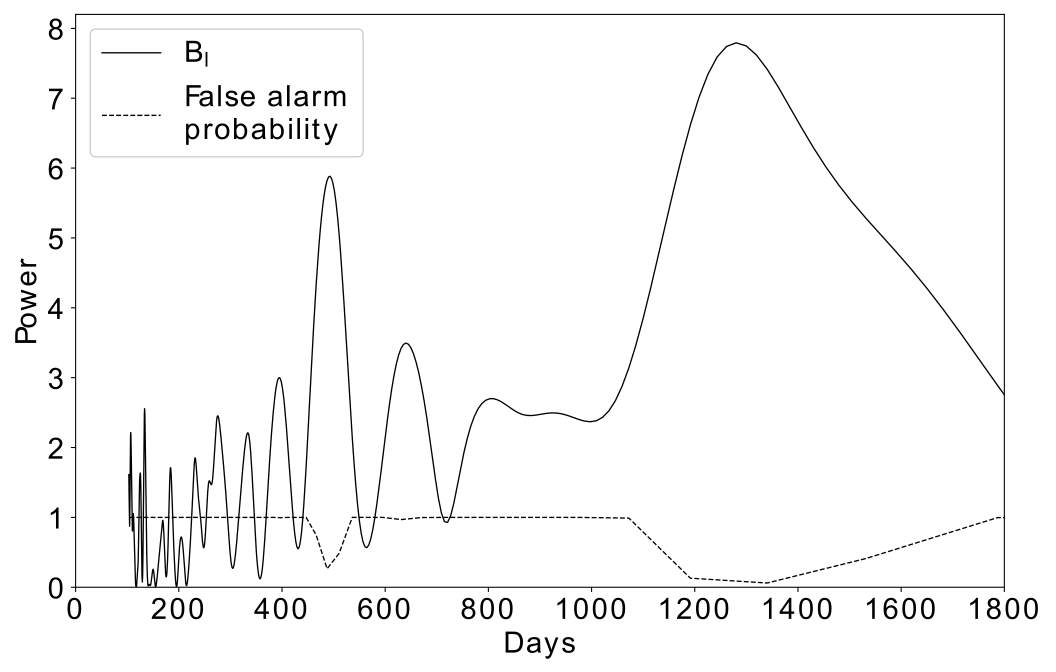}
    \includegraphics[width=0.48\textwidth]{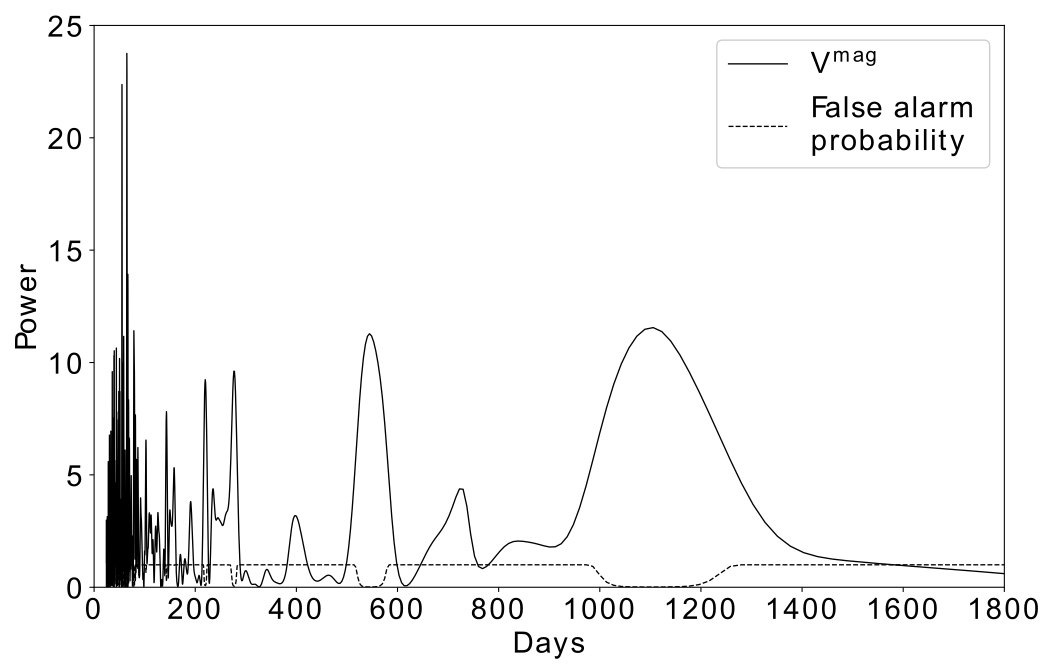}
    \caption{Lomb--Scargle period search for the spectral activity indicators (upper left panel), radial velocity (upper right), $B_l$ (bottom left), and the photometric light curve (bottom right). The false-alarm probability for each of them is designated with dashed lines.}
    \label{fig:lomb_scargle_indicators}
\end{figure}

\noindent We applied the Lomb--Scargle method \citep{lomb1976, scargle1982} to search for a period in the $B_l$, radial velocity, activity indicator, and photometric data. The results are presented in Figure~\ref{fig:lomb_scargle_indicators}. The identified periods and their false-alarm probability (FAP) are summarized in Table~\ref{tab:periods}.\\

\noindent We identified different periods for $B_l$ and for the activity indicators and radial velocity. For $B_l$, we found two periods of 1280 and 493 days. The latter value, while presenting a higher false-alarm probability, is very close to the LSP of 480 days deduced from previous photometric studies \citep{Percy2016,Tabur2009,Percy2008}. On the other hand, the first period is longer than the upper limit for the rotation period (909 days) and cannot thus be explained by rotational modulation. For the activity indicators and radial velocity, a clear period of about 704 days was identified. Taking into account that this period is not identified in the magnetic field and that these periods are not multiples of each other, we assume that this structure is not of magnetic origin. It is possible that this period results from an atmospheric feature such as a large vortex, as described in \cite{kapyla2011}, who proposed that features like this could appear in the atmospheres of fast-rotating and convective late-type stars and might contribute to rotationally modulated variations in the brightness and spectrum of the star.\\

\begin{table}
\caption{Significant Lomb--Scargle periods longer than 200 days for RZ~Ari.}
\begin{center}
\begin{tabular}{llll}
\hline
Indicator& Period & accuracy& FAP \\
      &  days    & interval&percent \\ \hline
$B_l$ & 1280 & 1243, 1319& 6.1 \\
      & 493 &487, 498 & 26.4 \\
S-index& 688 & 668, 707 & 2.0 \\
Ca IRT-index  & 717 & 696, 738& 0 \\
$H_\alpha$-index& 707 & 687, 728& 0.8  \\
$v_{rad}$ & 561 & 551, 570 & 0.7  \\
      & 717 &705, 733 & 0 \\
V-band phot.& 1105& 1089, 1122& 0.7  \\ 
            &544& 540, 548& 1.0 \\ \hline
\end{tabular}
\end{center}
\label{tab:periods}
\end{table}

\noindent From the LSD profiles obtained with the restricted line mask ($\chi_{\rm thr} > 2.5$~eV), we evaluated the heliocentric radial velocity (HRV) of the star by fitting the LSD Stokes~$I$ profiles with a Gaussian and taking the position of its peak as the HRV. We then performed a period-search analysis using the Lomb-Scargle method on the resulting HRV values. We obtained two periods of 561 and 717 days, which are listed in detail in Table~\ref{tab:periods}.\\

\noindent Finally, the Lomb-Scargle peridogram of the AAVSO photometric data indicates significant periods (FAP lower than 5\%) of 544.6 days and 1105.6 days. However, the first period is about two times shorter than the second and might be a harmonic of the main period because of the nonsinusoidal shape of the curves.      

\subsection{Zeeman-Doppler period search for RZ~Ari}
\label{sec:period-zdi}
\begin{figure}[h]
    \centering
    \includegraphics[width=\columnwidth]{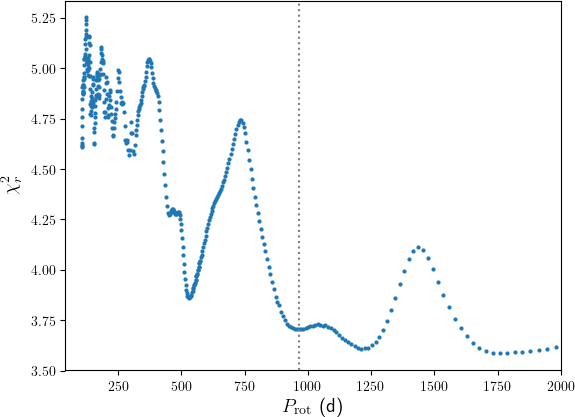}
    \caption{ZDI period search for the Stokes~$V$ profiles dataset for RZ~Ari. The calculated upper limit for the rotation period is plotted as the vertical dotted line.}
    \label{fig:zdi-period}
\end{figure}

In order to complement the period search based on the longitudinal magnetic field performed in the previous section, we attempted to model the time series of Stokes~$V$ signatures of RZ~Ari (obtained with the full-line mask) with ZDI (Donati et al. 1997). With this approach, we can use the full information available in the Stokes~$V$ LSD profiles, whereas longitudinal field measurements only use the first moment of these profiles. However, the ZDI method assumes that the variability observed in the spectral line profiles is only due to rotational modulation associated with steady features at the stellar surface. This assumption does not hold for RZ~Ari because \textit{(i)} the magnetic field evolves on a timescale comparable to the stellar rotation period, and \textit{(ii)} the star pulsates. In order to mitigate these effects, we only tried to model 28 polarimetric sequences collected between September 2010 and December 2015, that is, during a time span of $1\,930$~days. A Zeeman signature is definitely detected in most of these Stokes~$V$ profiles, and a clear variation in $B_l$ is observed during this time span. In addition, each Stokes~$I$ and $V$ profile is shifted to its rest frame velocity (determined with a Gaussian fit to the Stokes~$I$ profile), and its equivalent width is normalised to the average value of the time series. After applying this preprocessing, starting from an initial reduced $\chi^2$ of $7.53$, it is possible to fit our time series of Stokes~$V$ LSD profiles with ZDI down to a reduced $\chi^2$ of $3.0$. In order to derive the rotation period that fits our dataset best, we proceeded in a similar way as \cite{petit2002}. For a range of rotation periods between 40~d and 2000~d, we derived the $\chi^2$ level at which the Stokes~$V$ spectra can be modeled with ZDI at a given information content (i.e. ,for a fixed averaged magnetic value). The result is shown in Fig.~\ref{fig:zdi-period}. The resulting curve features four main minima. Following Petit et al. (2002), we locally fit a parabola to the $\chi^2_r$ curve to derive the corresponding periods and uncertainties. The main minima are located at $P_{\rm rot} = 530 \pm 2$~d, $967 \pm 11$~d, $1\,271\pm11$~d, and $1\,760 \pm 46$~d, and those at 967 and 1271~d are part of a broader minimum. Except for the longest minimum, these period values are close to those derived from longitudinal field measurements. The minimum at 530~d is also close to the photometric period. However, the $\sim 700$~d period recovered from radial velocity and activity indicators is clearly excluded as a possible rotation period by the ZDI analysis. The only rotation period value that is fully compatible with our determinations of $v\sin i$ and the radius is the period at 530~d, while the periods at $967 \pm 11$~d and  $1\,271\pm11$~d would suggest a slight mismatch with our $v\sin i$ and radius determination and a star seen nearly equator-on.

\noindent The fit to the Stokes~$V$ time series displays clear discrepancies that agree with the $\chi^2_r$ level. This confirms that the magnetic field varies significantly during one stellar rotation. The corresponding ZDI map (for the inclination angle of 40 degrees mentioned in Appendix A.1 for $P_{\rm rot} = 530$~days) can only be considered as indicative of the surface magnetic topology of RZ~Ari. It is shown in Appendix~\ref{sec:app-zdi} along with the corresponding fit to Stokes~$V$ LSD profiles for the same period. It is predominantly poloidal (90~\% of the magnetic energy on the map shown, and $>80~\%$ for the other rotation periods) and rather axisymmetric (75~\% of the magnetic energy on the map shown, and $>60~\%$ for the other rotation periods).

\subsection{Asymmetry in the intensity profiles}
\label{sec:asymmetry}

The Stokes~$I$ LSD profiles of RZ~Ari obtained with the full-line mask have symmetric shapes, but those obtained with the restricted mask sometimes display significant asymmetry between the blue and red wings of the profile. This has already been reported by \cite{Georgiev2020}, who suspected that high excitation potential lines (which are formed deeper in the atmosphere) are sensitive to downward motions, and this might be the reason for the distortion of their profiles in the red wing, where a dip is sometimes observed. The authors also suggested that no such asymmetry is observed when a full-line mask is used because the lines with lower excitation potential cancel this effect out.\\
The asymmetry in the Stokes~$I$ profiles varies with time. It is sometimes present to a different degree, and is absent altogether at other times. Significant asymmetry is present in the profiles of all observations in the following time intervals: November 16, 2011 to July 18, 2012; 2012/09/04 to 2012/11/12; the single date 2013/11/07; 5 August, 2016 to September 1, 2016; October 29, 2016 to December 20, 2016; the single date October 30, 2017; and January 23, 2018 to November 16, 2018.\\
This variability does not seem to correlate with any of the indicators of magnetic activity that we study. Significant asymmetry in the Stokes~$I$ profiles is observed both when a strong longitudinal magnetic field is measured (e.g., October 16, 2011) and when no signal in Stokes~$V$ is present at all (e.g., October 29, 2016 and October 22, 2018). On the other hand, some observations show a strong longitudinal magnetic field and very symmetrical Stokes~$I$ profiles (e.g., September 26, 2011 and October 31, 2015). The variability of this asymmetry does not correlate with the variability of any of the spectral activity indicators either. It is observed both when the individual indicators are high and when they are low. It therefore appears that the motions in the deep layers of the stellar atmosphere, which cause this observed asymmetry, do not affect the surface magnetic field.

\section{Linear polarization}\label{sec:linear}

\begin{figure*}[h]
    \centering
    \begin{minipage}[b]{0.775\textwidth}
        \includegraphics[width=\textwidth]{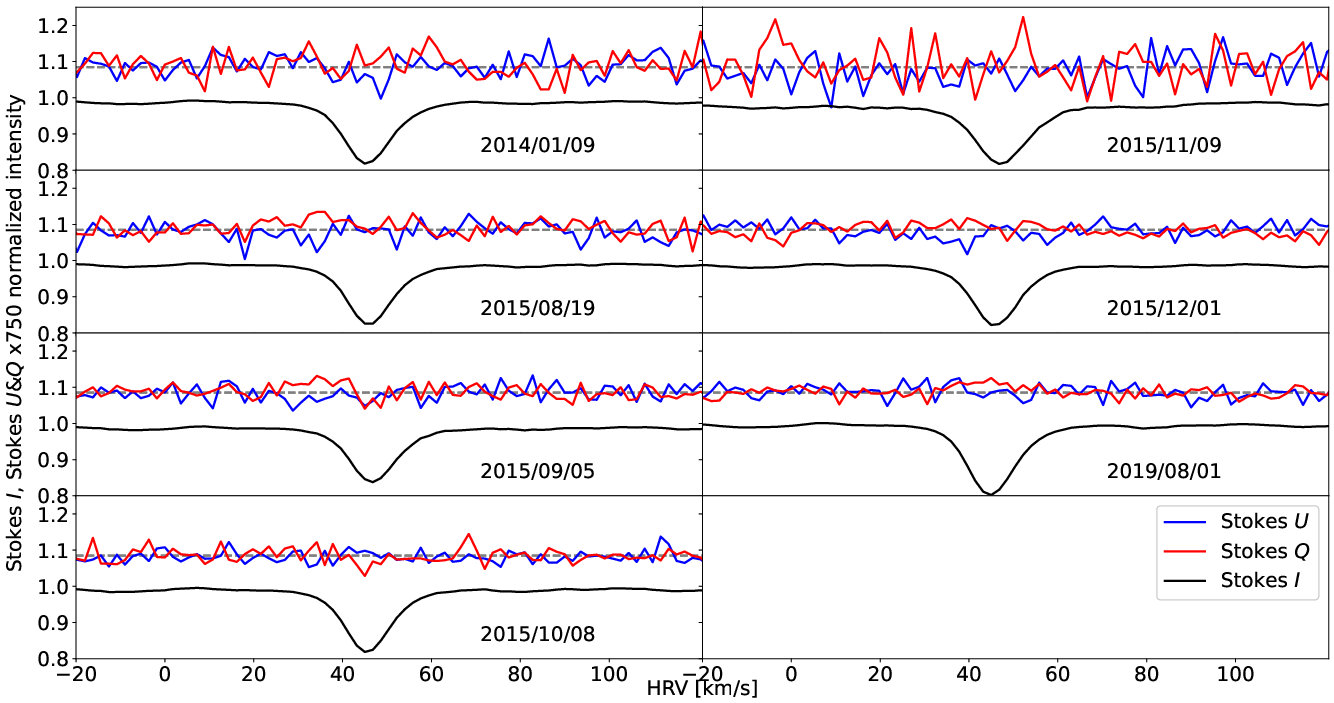}
    \end{minipage}\hfill
    \begin{minipage}[b]{0.205\textwidth}
        \caption{LSD profiles of RZ~Ari in linear polarization. Each panel represents a particular observation date. For each observation, the LSD Stokes~$I$ (in black) profile is shown, along with the vertically shifted and amplified Stokes~$U$ (in blue) and $Q$ (in red) profiles. No detection of polarized signatures is found in either Stokes parameter for any observation.}
    \end{minipage}
    \label{fig:linear}
   \end{figure*}

 We obtained seven observations of RZ~Ari in linear polarization, the LSD profiles of which are displayed in Figure~\ref{fig:linear}. No detection of polarized signatures is found for any of the observations, neither in Stokes~$U$ nor in Stokes~$Q$. This result differs from what was obtained for the supergiant Betelgeuse \citep{auriere2016} where the linear polarization is explained by giant convective cells, or in the Mira-type giant  $\chi$~Cyg  \citep{lebre2014}, where shock waves propagate in the atmosphere during certain phases of the pulsation cycle. No linear polarization signal was detected in the M giant $\beta$~Peg either \citep{Georgiev2020betpeg}. However, our observations in linear polarization are only few in number and were mostly collected during 2015. Hence, we cannot use this result as firm evidence for the absence of giant convective cells in RZ~Ari. It is also possible that the size of the large convective structures in RZ~Ari is smaller than in the supergiant Betelgeuse and does not result in
linear polarization to this degree.\\

\noindent On the other hand, we have to point out that in 2015, the linear polarization observations were performed during the same nights as the circular polarization observations. Magnetic field of different strengths was detected then. We can conclude that at least for this period, when we find no linear polarization (and hence, no giant convective cells) we should not expect that a local dynamo contributes significantly to the magnetic field.

\section{Discussion}\label{sec:discussion}

\subsection{The role of pulsations}

\begin{figure}[h]
   \centering
   \includegraphics[width=0.95\columnwidth, angle=0]{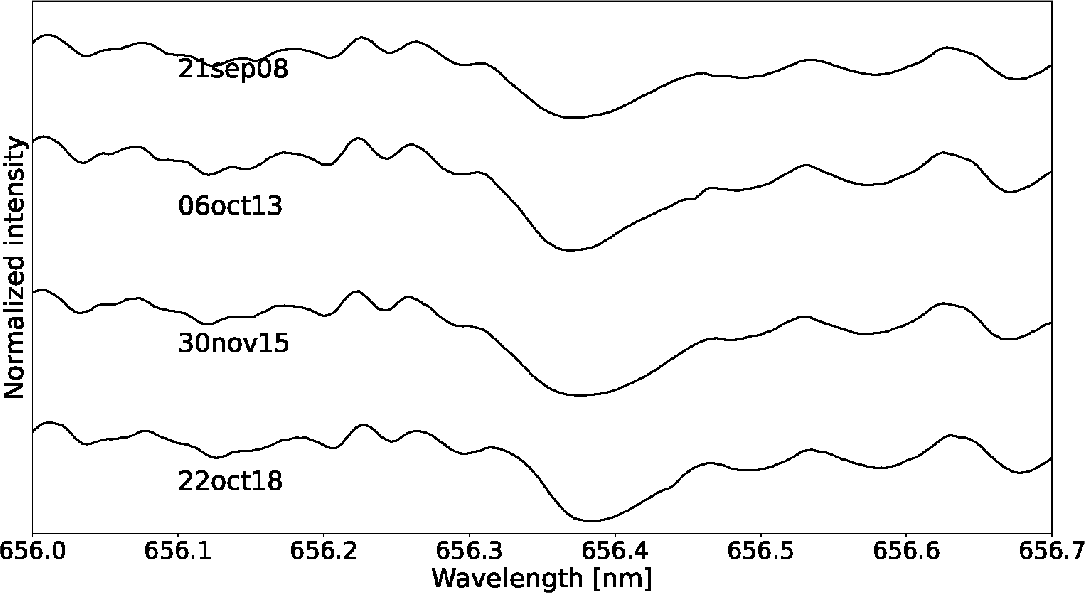}
     \caption{Example profiles of the $H_\alpha$ line in RZ~Ari. Spectra taken on different dates are shifted vertically for display purposes. No variability typical in the presence of an atmospheric shock wave is observed.}
         \label{fig:halpha}
   \end{figure}

RZ~Ari is a semiregular pulsating star, with a pulsation period of 56.5 days and an amplitude of its magnitude variation of 0.5~mag in V filter (from the General Catalogue of Variable Stars, \citealt{samus2017}). Taking into account the similarity to other long-period variable pulsating AGB stars such as Mira stars, we investigated the collected spectra to search for evidence of shock waves propagating throughout the stellar atmosphere. However, no typical feature related to strong radiative shock waves in the atmosphere of RZ~Ari could be found, such as the (periodic) splitting of atomic lines, emission within the hydrogen line profiles, or strong variability in these line profiles. The spectroscopic behavior of the $H_\alpha$ line is illustrated in Fig.~\ref{fig:halpha} for some dates.
We conclude that the behavior of the line profiles and that of $v_{\rm rad}$ shows no evidence of shocks at any phase of the pulsation period we covered with our Narval observations. Moreover, we simultaneously studied the photometric variability together with the variability of the magnetic field and activity indicators (e.g., see Fig.~\ref{fig:bl_ind_vrad}). On this basis, we conclude that possible shock waves (if they exist in RZ~Ari) cannot explain the magnetic field behavior in this giant, as was proposed for other pulsating cool evolved stars, such as the Mira-type star $\chi$Cyg  \citep{lebre2014} or the RV~Tauri star R~Sct \citep{Sabin2015,georgiev_thesis,Georgiev2023}. submitted). 
The mechanism for magnetic field generation in RZ~Ari is thus very likely different from the compression of a weak magnetic field by the passage of shock waves.\\

\noindent In addition, we cannot find evidence of shock waves in two other magnetic and pulsating early AGB stars, EK~Boo and $\beta$ Peg \citep{georgiev_thesis}. Hence, the spectral observations show no strong shocks in the studied semiregular pulsating stars at this evolutionary stage. It is rather possible that strong shocks appear later in the evolution and at a lower surface gravity of the star.

\subsection{Are there giant convective cells in RZ~Ari?}

Giant convective cells were recently detected by means of interferometry at the surface of the 1.5 M$_{\odot}$ AGB star $\pi^1$~Gru \citep{paladini2018}. Giant convective cells also exist on the supergiant Betelgeuse, and a local dynamo causes the detected magnetic field in it \citep{auriere2010, mathias2018}. In the past, these patterns in AGB and red supergiants were also predicted by numerical simulations of convection in \cite{Woodward2003} and \cite{Dorch2004} respectively. Giant convective cells also appear in RGB simulations, as shown by \cite{BrunPalacios2009}. The questions we try to answer is whether large convective cells like this exist in RZ~Ari, and if a local dynamo can explain the detected magnetic field and its variability there. We recall that the 1280-day magnetic field period is longer than the rotational period and is consistent with the lifetime predictions for large convective cells of 1000 days or more by \cite{freytag2017} and \cite{chiavassa2018}.

\noindent The different timescales of the convective structures (large convective cells and smaller granules) and their instability could explain why we observe periods of magnetic field detection and periods of nondetection in the M giants we study, including RZ~Ari. According to the models by Freytag et al. (2017), the lifetimes of the giant convective cells in AGB stars are about 1000 days, but for the smaller granules, the lifetime is one month and even less.

\noindent The LSP is also present in the magnetic field variability. We identified a period of 493 days, but with a high FAP. This FAP indicates a possible instability of the period. It is also not far from the 530-day period found by means of ZDI that is considered to be the rotation period of the star, but it is outside its accuracy interval. On the other hand, the 1280-day period (also found in $B_{l}$ variations) is very likely related to the lifetime of the convective structures. Similar long periods were also found by means of a ZDI period search and in the photometric data. On the other hand, we were unable to detect linear polarization, which is indicative for giant convective cells. However, we lack regular observations, and our few observations were mostly made in 2015. 
Further studies with the CHARA array to detect large convective structures, and later monitor them, will be very valuable.

\subsection{Dynamo considerations}

For an effective $\alpha$-$\omega$ dynamo operation, a $Ro$ lower than 1 is required. For RZ~Ari, if we assume a rotation period of about 530 days (upper limit of 909 days) and $\tau_{\rm conv}(H_p/2)$ less than 100 days, we obtain that $Ro$ is much more than 5, and therefore, an efficient $\alpha$-$\omega$ dynamo is unlikely there. Even when we consider $\tau_{\rm conv \ max}$, the longest convective turnover time in the convective envelope, which is about 37 days for a 1.5 M$_\odot$ star in the AGB stage, $Ro$ is about 14. Hence, an $\alpha-\omega$ dynamo operation is  very unlikely for RZ~Ari. Other types of dynamo are not excluded from operating there, however, such as  an $\alpha^2-\omega$ dynamo \citep{charbonneau2013}, which works in the case of larger $Ro$ in AGB stars \citep{soker2000}, turbulent dynamo \citep{soker_zoabi2002}, or a local dynamo as in the supergiant Betelgeuse \citep{auriere2010}.

\noindent The rotation of M giants has not been addressed specifically, and only scarce and heterogeneous sets of v$\sin i$ can be found in the literature. We refer to \cite{Zamanov2008}, who compiled published data and their own determinations of $v\sin i$ for 53 M-type field giants, and found a mean $v\sin i$ of 4.8~km\,s$^{-1}$ for their sample, covering spectral types M0III to M6III. However, macroturbulence is not taken into account in the determination of these values, and these values (and the mean value) should therefore be lower.
On the other hand, according to the models by \cite{privitera2016a}, the expected equatorial rotational velocity for stars with $\log g$ of 0.5 is about 2~km\,s$^{-1}$. 
In this context, RZ~Ari, for which we determined for $v\sin i$ a value of 6.0 km\,s$^{-1}$  \citep{Georgiev2020}, taking the macroturbulence into account (previous value by \cite{Zamanov2008} was 9.5 km\,s$^{-1}$ ) exhibits a higher surface rotation velocity compared to most M-type giant stars. We note that recent studies also assigned velocities higher than 5 km\,s$^{-1}$ to other M5-M6 III giants : IRAS 12556–7731 with $v\sin i$ = 8 km\,s$^{-1}$ \citep{Alcala2011}, or KIC 10526137 with $v\sin i$ = 10.3 km\,s$^{-1}$ \citep{Frasca2022}. However, these stars are assumed to be engulfed by a planet (see also subsection 6.4). From the theoretical side, these giants are the descendants of A to early-F main-sequence stars. The self-consistent evolution of their surface rotation rate is not fully understood. Stellar evolution models including rotation usually predict very slow rotation rates for these giants \citep[e.g. lower than 1 km\,s$^{-1}$, see][]{privitera2016a}, but these models are known to be incomplete and not to be able to account for the observed evolution of the surface (and core) rotation of low-mass stars.

\noindent With this faster rotation in RZ~Ari, we cannot exclude a contribution of the rotation to the dynamo operation. In addition, taking into account that the turbulent dynamo is not expected to produce surface fields stronger than 1 G in giants \citep{soker_zoabi2002}, the $\alpha^2-\omega$ dynamo remains a possible option. In this, the $\alpha$-effect and differential rotation shear contribute to the toroidal field production, according to the predictions for $\alpha^2-\omega$ dynamo operation in AGB stars by \cite{soker2000}. It is possible that in RZ~Ari, we observe the specifics of such a dynamo operation in a fairly evolved star, where the stellar structure and atmospheric dynamics are very different than in the main sequence stars.

\subsection{A scenario for potential planet engulfment}

A magnetic field is detected in RZ-Ari, this star is outside the magnetic strips in which $\alpha-\omega$ is thought to operate, and it is rotating fairly fast. Together with the chemical composition of RZ-Ari, in particular its lithium abundance, another possibility to explain its dynamo could be that it is triggered by planet engulfment.

\noindent RZ~Ari might be a less extreme version of IRAS 12556–7731 \citep{Alcala2011}, where planet engulfment is also suspected to cause its relatively fast rotation\footnote{Usually, for the RGB stars, the limit between slow and fast rotating giants is associated with v$\sin i$ between 5 to 10 km\,s$^{-1}$ \citep[see e.g.][for discussion]{Carlberg2012}. However, it is not studied  for the AGB stars.} and high lithium content. However, its magnetic field and activity are not studied. According to \cite{Carlberg2012}, planet engulfment in giant stars could be advocated to explain lithium enrichment for stars with $\log\varepsilon({\rm Li})$ $\leq$ 2.2, which is the case of RZ-Ari. Furthermore, \cite{privitera2016c,privitera2016a,privitera2016b} studied the impact of planet engulfment during the RGB evolution on the rotation and magnetic fields of stars with initial masses between 1.5 and 2.5 M$_\odot$ at solar metallicity. They focused on the case of planets with various orbital distances that were less massive than 15 M$_{Jupiter}$ , which would be fully destroyed by the engulfment so that the integrity of their orbital angular momentum should be transferred to the star. Focusing on the 1.5 M$_\odot$ case, with an initial rotation $\Omega/\Omega_{\rm ini} = 0.5$, which can be used as representative of RZ-Ari, they predict that such a RGB star with $\log g\approx 0.3$ could exhibit a surface velocity of about 6 km\,s$^{-1}$ after the engulfment of a planet of more than 10 M$_{\rm Jupiter}$ earlier during its red giant branch ascent (meaning several 10$^7$ yr earlier in the evolution). They indicated that no specific variation of $^{12}C/^{13}C$ is expected following the engulfment, and that lithium enrichment could occur, but would be difficult to use as a tracer of engulfment per se. \cite{privitera2016c} also evaluated how an engulfment like this might trigger high magnetic fields at the surface of giants. Their models showed that the large increase in surface velocity at engulfment probably results in a lowering of the Rossby number and might trigger a $\alpha-\omega$ dynamo in the convective envelope, which would generate a magnetic field. In their planet engulfment scenario, the maximum longitudinal magnetic field expected immediately following the engulfment should be about 20 Gauss. It is then predicted to decay as the Rossby number increases again (and the $\alpha-\omega$ dynamo can no longer be sustained) while the star continues to evolve to the RGB tip over several million years to reach the Gauss level or less near the RGB tip, which is compatible with the mean behavior of our RZ~Ari observations. \\ We finally note that these studies correspond to RGB stars. When we consider that RZ-Ari is on the AGB (see \S~\ref{sec:evol}), as on its way to its current evolutionary point, it would have already reached the same radius, $\log$g, and luminosity (because the stellar tracks in this region of the HR and Kiehl diagrams are degenerate), the planet engulfment scenario would be more difficult to advocate, but not impossible. In recent works on planetary systems dynamics along stellar evolution up to the tip AGB,  it was shown that the possible survival of planets during the evolution after RGB probably is a result of the interplay of the tidal forces and the stellar wind that expands the planetary orbits \citep{mustill2012}. Observations indeed found some close orbiting planets near horizontal branch stars that apparently survived the RGB phase \citep{Silvotti2007, Charpinet2011, Setiawan2010}. Hence, the region closer to the star can be repopulated at the end of the RGB. \cite{Trifonov2022} performed a numerical study of the evolution of a planetary system around a star similar to RZ-Ari. They reported that if a planet is more distant than a certain limit (depending on the mass of the star, the mass of the planet, and the architecture of the planetary system) it should migrate to a larger orbit before the AGB phase and could be engulfed later, when the star is ascending the AGB. The fact that planets exist even around white dwarfs supports the model prediction that some planets may survive not only the RGB, but also the AGB \citep{Veras2016}. While the scenario of planetary engulfment on the AGB is not as obvious as the planet engulfment on the RGB, we cannot exclude it as a possible interpretation of RZ~Ari rotation, surface lithium abundance, and magnetic activity, regardless of whether RZ-Ari is at the RGB tip or an early AGB star.\\

\section{Conclusions}

   \begin{enumerate}
      \item The M6 giant RZ~Ari was observed with Narval in the period September 2010 -- August 2019 during 56 nights.
A magnetic field  was definitely detected in the photosphere of this giant, and 
its longitudinal component $B_l$ was mostly about a few Gauss, but in October and November 2011, it was more than 10 Gauss. The activity indicators are variable.  The periods of variation were derived for the magnetic field, activity indicators, radial velocity, and for the photometric light curve. Periods longer than 1100 days are found for the magnetic field and photometric light curve. This long-period variability was also confirmed by the ZDI period search. The so-called LSP reported in the literature was also identified in the magnetic field variability. It is not so different from the period of 530 days found by means of ZDI and photometry period searches, and it is considered to be the rotational period of the star. However, not all detected periods in the magnetic field and photometric light curve are evident in the activity indicators, where a period of about 704 days is found. The longer period of 704 days does not to appear to be of magnetic origin and might be a large vortex in the extended atmosphere of the giant, as predicted by the theory, or some other convective structure.
\item On the basis of precise data for its $v\sin i$, radius, effective temperature, and luminosity, the evolutionary status and mass of the star were determined precisely, as well as the upper limit for the rotational period. The star has a mass of 1.5 M$_{\odot}$ and is either at the tip of the RGB or on the AGB. The estimated upper limit for the rotation period is 909 days. Hence, the identified period of about 530 days by means our observations may in fact correspond to the rotation period of the star.
\item We also scrutinised the line profiles of hydrogen (the $H_\alpha$ line) and metallic elements and found no strong shock waves propagating throughout the atmosphere of the star. With this result, we conclude that magnetic field compression due to atmospheric dynamics involving shock-wave propagation is not a likely explanation for the magnetic field we detect in RZ~Ari.
\item A dynamo of the $\alpha-\omega$ type is also unlikely 
to operate there because of the high $R_o$ number of the star. Moreover, we found that RZ~Ari is a fast-rotating and  Li--rich star and is also situated outside the second magnetic strip, where, according to the models, $R_o$ is lower than 1. One possibility is the $\alpha^2-\omega$ dynamo, which works with $R_o$ values higher than 1 to operate there. 
In general, the driving mechanisms for the 
$\alpha^2-\omega$ dynamo operation on the giant branches could be angular momentum dredge-up 
from the interior, binary merging, or planet engulfment. The high lithium content and fast rotation favor the later. A turbulent dynamo appears to be challenged by the field strength measured for RZ~Ari. According to the model predictions, a turbulent dynamo yields a magnetic field of about 1 Gauss or less at the surface of the giant stars.

At the moment, we have no direct evidence of large convective cells at the surface of this M giant. No linear polarization is detected for it during our spectropolarimetric observations, which were mostly made in 2015. However, a magnetic field is detected then. Hence, we do not expect a local dynamo to contribute significantly to the magnetic field at least for this time interval. On the other hand, the periods longer than 1000 days we found in the magnetic field variability and in the photometric light curve cannot be explained by the rotational modulation of the star. However, they are consistent with the lifetime of large convective cells predicted by the models. Further interferometric study in this direction is highly desirable. 
      
   \end{enumerate}

\begin{acknowledgements}
      We thank the anonymous referee for the useful questions that led to improvement of the paper text. We thank the TBL team for their service observing support. Our observations were funded by OPTICON, Bulgarian NSF projects DSAB 02/3 and DN 18/2 and "Programme National de Physique Stellaire" (PNPS) of CNRS/INSU co-funded by CEA and CNES. The authors also acknowledge use of the AAVSO database. A.P. thanks to Prof. Corinne Charbonnel for insights and macros, and Dr. Antonio Frasca for providing his data concerning the rotation, lithium abundance and stellar parameters of Kepler M giants. R.K.-A. is thankful to Drs. Thomas Hackman, Berndt Freytag  and Trifon Trifonov for the useful discussions. We further acknowledge use of the Gaia DR2 database, the VizieR catalogue access tool (CDS, Strasbourg, France) and of the VALD data base in Vienna. F.B. and C.A. acknowledge funding from NSF AAG award 1814777. NAD acknowledges Funda\c c\~ao de Amparo \`a Pesquisa do Estado do Rio de Janeiro - FAPERJ, Rio de Janeiro, Brazil,  for grant E-26/203.847/2022. R.K.-A., R.B., S.G., S.Ts., A.L. and A.P. acknowledge partial financial support by the Bulgarian NSF under contract DN 18/2.
\end{acknowledgements}

\bibliographystyle{aa}
\bibliography{main}

\onecolumn
\appendix 
\section{Additional table}
\label{sec:app-table}
\begin{table*}[h]
\centering
{\tiny\renewcommand{\arraystretch}{1.0}
\begin{tabular}{c c c c c c c c c c}
\hline\hline                 
Date     &HJD      & Exposure time   &Detection     & B$_{\rm l}$ [G]& $\sigma$ [G]&S-index & H$\alpha$-index & CaIRT-index & $v_{\textrm{rad}}$ [km\,s$^{-1}$]\\ \hline
2010/09/05 & 5445 & 10x600s & ND & N/A & N/A & 0.206 & 0.407 & 0.734 & 47.6\\
2010/09/21 & 5461 & 16x400s & DD & 3.47 & 0.36 & 0.148 & 0.382 & 0.685 & 47.7\\
2010/10/13 & 5483 & 16x400s & DD & 1.17 & 0.28 & 0.153 & 0.379 & 0.669 & 47.4\\
2011/01/22 & 5584 & 16x400s & DD & 1.82 & 0.53 & 0.163 & 0.400 & 0.727 & 48.2\\
2011/01/27 & 5589 & 4x400s & ND & N/A & N/A & 0.146 & 0.399 & 0.696 & 47.9\\
2011/02/04 & 5597 & 16x400s & DD & 0.99 & 0.33 & 0.150 & 0.388 & 0.660 & 46.8\\
2011/09/26 & 5831 & 16x400s & DD & 6.68 & 0.43 & 0.274 & 0.327 & 0.614 & 45.0\\
2011/10/16 & 5851 & 16x400s & DD & 14.01 & 0.34 & 0.223 & 0.322 & 0.657 & 46.4\\
2011/11/23 & 5889 & 4x400s & DD & 11.06 & 0.70 & 0.232 & 0.311 & 0.638 & 47.0\\
2011/11/24 & 5890 & 12x400s & DD & 10.45 & 0.50 & 0.232 & 0.311 & 0.638 & 45.9\\
2012/01/10 & 5937 & 8x400s & DD & 11.17 & 0.47 & 0.199 & 0.313 & 0.592 & 45.6\\
2012/01/11 & 5938 & 8x400s & DD & 10.73 & 0.52 & 0.206 & 0.313 & 0.598 & 45.8\\
2012/07/16 & 6125 & 2x400s & N/A & N/A & N/A & 0.218 & 0.326 & 0.633 & 46.8\\
2012/07/17 & 6126 & 4x400s & N/A & N/A & N/A & 0.218 & 0.326 & 0.633 & 46.8\\
2012/07/18 & 6127 & 2x400s & N/A & N/A & N/A & 0.218 & 0.326 & 0.633 & 46.9\\
2012/08/16 & 6156 & 5x400s & N/A & N/A & N/A & 0.193 & 0.335 & 0.677 & 48.3\\
2012/08/17 & 6157 & 3x400s & N/A & N/A & N/A & 0.193 & 0.335 & 0.677 & 48.3\\
2012/09/04 & 6175 & 5x400s & N/A & N/A & N/A & 0.195 & 0.328 & 0.631 & 47.4\\
2012/09/05 & 6176 & 3x400s & N/A & N/A & N/A & 0.191 & 0.326 & 0.629 & 47.3\\
2012/10/04 & 6205 & 8x400s & DD & 2.58 & 0.49 & 0.188 & 0.339 & 0.703 & 47.7\\
2012/11/12 & 6244 & 8x400s & DD & 5.30 & 0.61 & 0.169 & 0.347 & 0.727 & 47.1\\
2013/01/11 & 6304 & 8x400s & DD & -0.43 & 0.53 & 0.163 & 0.354 & 0.698 & 46.4\\
2013/07/08 & 6482 & 2x400s & ND & N/A & N/A & 0.195 & 0.388 & 0.727 & 47.9\\
2013/08/05 & 6510 & 3x400s & ND & N/A & N/A & 0.202 & 0.370 & 0.611 & 45.1\\
2013/09/02 & 6538 & 4x280s & DD & -3.63 & 0.78 & 0.159 & 0.388 & 0.684 & 46.8\\
2013/10/06 & 6572 & 4x400s & ND & N/A & N/A & 0.260 & 0.415 & 0.728 & 46.0\\
2013/11/07 & 6604 & 4x400s & ND & N/A & N/A & 0.215 & 0.388 & 0.687 & 46.4\\
2013/12/03 & 6630 & 4x400s & ND & N/A & N/A & 0.247 & 0.366 & 0.644 & 45.2\\
2014/01/09 & 6667 & 4x400s & DD & -0.97 & 0.58 & 0.214 & 0.365 & 0.607 & 45.5\\
2015/08/19 & 7254 & 8x400s & DD & 3.07 & 0.62 & 0.203 & 0.382 & 0.618 & 46.0\\
2015/09/05 & 7271 & 8x400s & DD & 4.31 & 0.61 & 0.209 & 0.383 & 0.635 & 46.7\\
2015/10/08 & 7304 & 8x400s & DD & 3.17 & 0.45 & 0.229 & 0.354 & 0.613 & 45.6\\
2015/10/31 & 7327 & 8x400s & DD & 4.78 & 0.56 & 0.218 & 0.358 & 0.674 & 47.1\\
2015/11/30 & 7357 & 8x400s & DD & 2.07 & 0.55 & 0.254 & 0.349 & 0.633 & 45.9\\
2015/12/18 & 7375 & 8x400s & DD & 1.79 & 0.53 & 0.222 & 0.359 & 0.663 & 47.4\\
2016/08/05 & 7606 & 8x400s & ND & N/A & N/A & 0.191 & 0.397 & 0.637 & 47.3\\
2016/09/01 & 7633 & 8x400s & ND & N/A & N/A & 0.175 & 0.407 & 0.639 & 46.5\\
2016/10/03 & 7665 & 8x400s & MD & -1.41 & 0.50 & 0.155 & 0.431 & 0.700 & 47.2\\
2016/10/29 & 7691 & 8x400s & ND & N/A & N/A & 0.176 & 0.425 & 0.740 & 47.8\\
2016/12/01 & 7724 & 8x400s & ND & N/A & N/A & 0.213 & 0.406 & 0.735 & 47.5\\
2016/12/20 & 7743 & 8x400s & ND & N/A & N/A & 0.215 & 0.408 & 0.745 & 47.9\\
2017/01/07 & 7761 & 8x400s & ND & N/A & N/A & 0.193 & 0.402 & 0.727 & 48.0\\
2017/02/16 & 7801 & 8x400s & ND & N/A & N/A & 0.208 & 0.396 & 0.708 & 46.8\\
2017/09/04 & 8001 & 8x400s & DD & -1.77 & 0.86 & 0.206 & 0.360 & 0.666 & 46.1\\
2017/10/06 & 8033 & 8x400s & DD & -0.11 & 0.57 & 0.246 & 0.336 & 0.623 & 44.8\\
2017/10/30 & 8057 & 8x400s & MD & -1.03 & 0.54 & 0.201 & 0.322 & 0.614 & 44.8\\
2017/11/23 & 8081 & 8x400s & ND & N/A & N/A & 0.244 & 0.346 & 0.662 & 46.1\\
2018/01/23 & 8142 & 8x400s & DD & -1.15 & 0.90 & 0.214 & 0.327 & 0.605 & 45.3\\
2018/09/18 & 8380 & 8x400s & ND & N/A & N/A & 0.196 & 0.395 & 0.739 & 47.7\\
2018/10/22 & 8414 & 7x400s & ND & N/A & N/A & 0.181 & 0.405 & 0.728 & 47.6\\
2018/11/16 & 8439 & 8x320s & ND & N/A & N/A & 0.193 & 0.406 & 0.727 & 47.6\\
2019/01/07 & 8491 & 8x400s & ND & N/A & N/A & 0.200 & 0.396 & 0.699 & 46.4\\
2019/01/26 & 8510 & 8x400s & ND & N/A & N/A & 0.283 & 0.371 & 0.737 & 46.7\\
2019/03/08 & 8551 & 2x400s & ND & N/A & N/A & 0.246 & 0.428 & 0.783 & 46.0\\
2019/03/11 & 8553 & 8x400s & ND & N/A & N/A & 0.213 & 0.388 & 0.714 & 45.9\\
2019/08/03 & 8699 & 8x400s & DD & -3.55 & 0.54 & 0.251 & 0.407 & 0.751 & 45.0\\
           &      &        &    &     &     &        &       &       &     \\
\hline
\end{tabular}}
\caption{Log of observations of RZ~Ari. Individual columns correspond to dates, heliocentric Julian dates (starting from 2\,450\,000), the total exposure time (number of spectropolarimetric sequences times the exposure time per sequence),  Zeeman detection (DD means a definitive detection, MD  means a marginal detection, and ND means no detection). N/A means no Zeeman signature examination due to the Fresnel rhomb misalignment, longitudinal magnetic field component, $B_l$, and its errors in G, Ca S-index, H$\alpha$- index, CaIRT-index, and radial velocity $v_{\rm rad}$ in km\,s$^{-1}$. Following the Nyquist theorem, we estimate the error of our $v_{\textrm{rad}}$ measurements to be 0.9 km\,s$^{-1}$.}
\label{tab:rzari_v}
\end{table*}
\newpage
\section{Tentative ZDI analysis of RZ~Ari}
\label{sec:app-zdi}

\begin{figure*}[h]
    \centering %
    \includegraphics[width=0.4\textwidth]{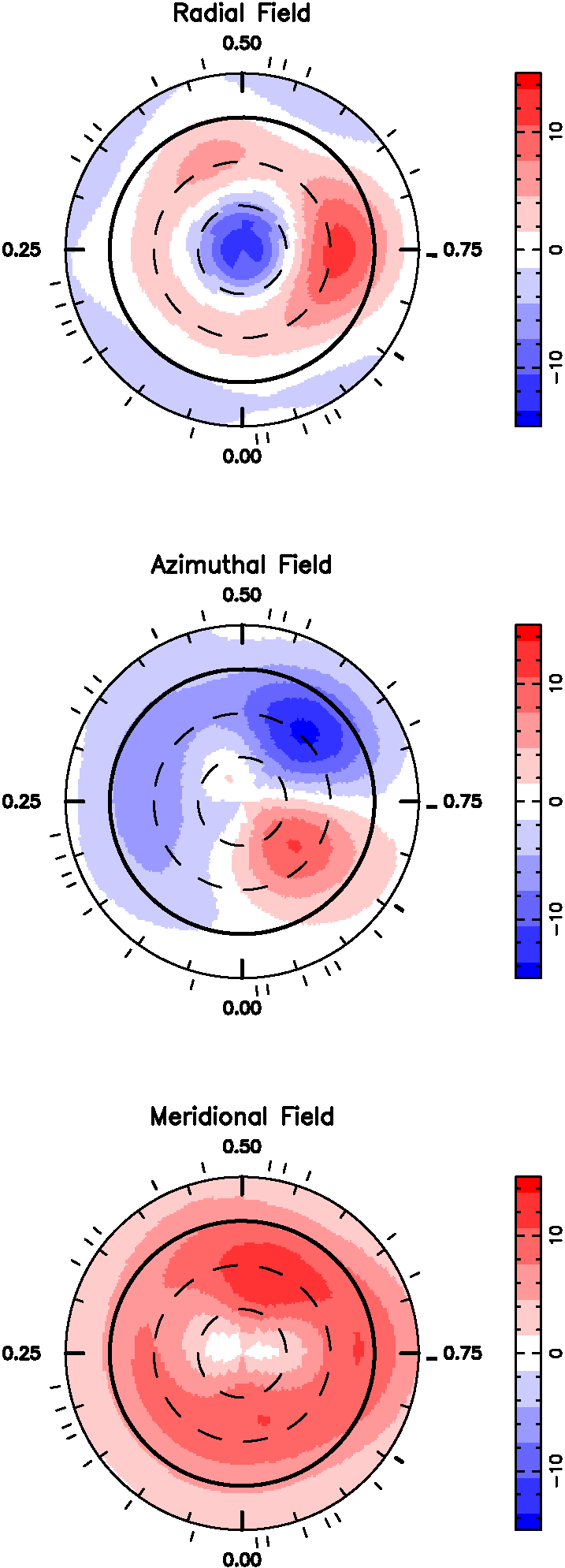}
    \caption{Tentative surface magnetic map of RZ~Ari reconstructed with ZDI; see section~\ref{sec:period-zdi}. The three components of the field in spherical coordinates are displayed from top to bottom (flux values are labeled in G). The star is shown in a flattened polar projection down to latitudes of -30$^{\circ}$, with the equator depicted as a bold circle and parallels as dashed circles. The radial ticks around each plot indicate phases of observations. For this reconstruction, we used $P_{\rm rot} = 530$~d and an inclination of the rotation axis with respect to the line of sight $i=40^{\circ}$}
    \label{fig:zdi-map}
\end{figure*}

\newpage 
\begin{figure}[h]
    \centering %
    \includegraphics[angle=0, width=\textwidth]{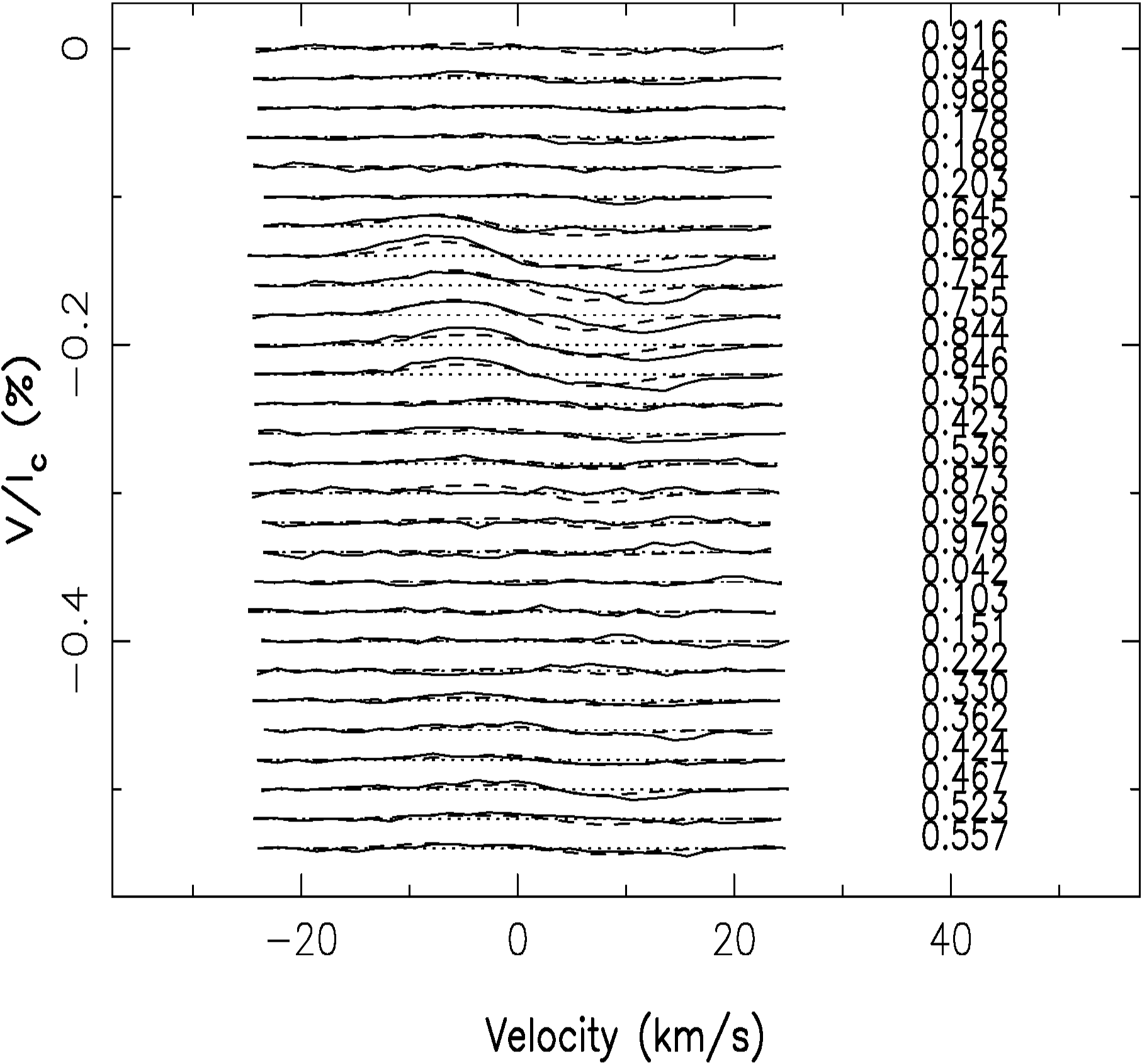}
    \caption{Fit to the Stokes~V LSD time series corresponding to the ZDI map of figure~\ref{fig:zdi-map}. Synthetic profiles corresponding to our magnetic models (dashed lines) are superimposed on the observed LSD profiles (solid lines). The fractional rotational phases of each observation are also mentioned at right-hand side of each profile. The successive profiles are shifted vertically for clarity purposes, and the associated reference levels (V = 0) are plotted as dotted lines.}
    \label{fig:zdi-fit}
\end{figure}

\end{document}